\documentclass[a4paper,11pt]{article}

\usepackage{jheppub} 

\usepackage[T1]{fontenc} 
\usepackage{CJK}
\usepackage{setspace}
\usepackage{float}
\def\be{\begin{equation}}
\def\ee{\end{equation}}
\def\bea{\begin{array}}
\def\eea{\end{array}}
\def\beqa{\begin{eqnarray}}
\def\eeqa{\end{eqnarray}}
\def\beqas{\begin{eqnarray*}}
\def\eeqas{\end{eqnarray*}}

\def\bp{\begin{picture}}
\def\ep{\end{picture}}
\def\bc{\begin{center}}
\def\ec{\end{center}}
\def\bfig{\begin{figure}}
\def\efig{\end{figure}}

\def\bit{\begin{itemize}}
\def\eit{\end{itemize}}
\def\nn{\nonumber}
\def\f{\frac}

\def\[{\left[}
\def\]{\right]}
\def\({\left(}
\def\){\right)}

\def\..{\left.}
\def\.{\right.}
\def\tl{\tilde}
\def\ra{\rightarrow}
\def\la{\leftarrow}

\def\tm{\times}

\def\da{\dagger}

\def\la{\lambda}

\def\al{\alpha}

\def\ep{\epsilon}

\def\pa{\partial}
\def\pr{\prime}

\title{Solving the muon g-2 anomaly in deflected AMSB with messenger-matter interactions}
\author[a]{Fei Wang,}
\author[b]{Wenyu Wang,}
\author[c,d]{Jin Min Yang}

\affiliation[a]{School of Physics, Zhengzhou University, Zhengzhou 450000, P. R. China}
\affiliation[b]{College of Applied Science, Beijing University of Technology, Beijing 100124, P. R. China}
\affiliation[c]{CAS Key Laboratory of Theoretical Physics, Institute of Theoretical Physics,
                Chinese Academy of Sciences, Beijing 100190, P. R. China}
\affiliation[d]{School of Physics, University of Chinese Academy of Sciences, Beijing 100049, P. R. China}

\emailAdd{feiwang@zzu.edu.cn}
\emailAdd{wywang@mail.itp.ac.cn}
\emailAdd{jmyang@itp.ac.cn}

\abstract{We proposed to introduce general messenger-matter interactions
in the deflected anomaly mediated SUSY breaking scenario to explain the $g_\mu-2$ anomaly.
Scenarios with complete or incomplete GUT multiplet messengers are discussed, respectively. The introduction of incomplete GUT mulitiplets can be advantageous in various aspects. We found that the $g_\mu-2$ anomaly can be solved in both scenarios under current
constraints including the gluino mass bounds, while the scenarios with incomplete GUT representation
messengers are more favored by the $g_\mu-2$ data.
We also found that the gluino is upper bounded by about 2.5 TeV (2.0 TeV)
in Scenario A and 3.0 TeV (2.7 TeV) in Scenario B if the generalized deflected AMSB scenarios are used
to fully account for the $g_\mu-2$ anomaly at $3\sigma$ ($2\sigma$) level.
Such a gluino should be accessible in the future LHC searches.
 Dark matter constraints, including DM relic density and direct detection bounds, favor the scenario B with incomplete GUT multiplets. Much of the allowed parameter space for the scenario B could be covered by the future DM direct detection experiments.

}

\begin{document}
\maketitle

\newpage
\section{Introduction}
Low energy supersymmetry (SUSY) is strongly motivated and
regarded as one of the most appealing candidates for
TeV-scale new physics beyond the Standard Model(SM).
SUSY can not only solve the gauge hierarchy problem of the SM,
but also elegantly explain the cosmic dark matter puzzle.
Besides, the gauge coupling unification, which can not be achieved
in the SM, can be successfully realized in the framework of SUSY.
Especially, the 125 GeV Higgs boson discovered by
the LHC \cite{ATLAS:higgs,CMS:higgs} lies miraculously in the narrow range
of $115-135$ GeV predicted by the Minimal Supersymmetric Standard Model (MSSM).

Although SUSY is an appealing extension of the SM, it seems
to have some tensions with the current LHC data. In particular, no evidences
of SUSY partners (sparticles) have been observed at the LHC.
Actually, the LHC data has already set stringent constraints
on sparticle masses \cite{CMSSM1,CMSSM2}
in simplified SUSY models, e.g., the gluino mass $m_{\tilde g} \gtrsim 1.9$
TeV for a massless lightest sparticle (LSP),  the lightest stop mass
$m_{\tl{t}_1}\gtrsim 850 $ GeV and even stronger bounds on the first two generations of squarks.
In fact, the LHC data agrees quite well with the SM predictions and no significant
deviations have been observed in flavor physics or electroweak precision measurements.
So far the only sizable deviation comes from the so-called anomalous magnetic moment of the muon
$a_\mu = (g_\mu-2)/2$ measured by the E821 experiment at the Brookhaven National
Laboratory \cite{muong-2:0208067}, which shows a 3.2$\sigma$ discrepancy from the SM prediction.
The SUSY explanation of this anomaly requires relatively light sleptons and electroweak gauginos.
If SUSY is indeed the new physics to explain all these experimental results, its spectrum must display
an intricate structure. Therefore, the origin of SUSY breaking and its mediation mechanism,
which determines the low energy SUSY spectrum, is a crucial issue.

There are many popular ways to mediate the SUSY breaking effects from the
hidden sector to the visible MSSM sector, such as 
the gravity mediation \cite{SUGRA}, the gauge mediation \cite{GMSB} and the anomaly mediation \cite{AMSB} SUSY breaking(AMSB) mechanisms.
Spectrum from the AMSB is insensitive to the ultraviolet(UV) theory \cite{deflect:RGE-invariance}
and automatically solves the SUSY flavor problem. Unfortunately,
the AMSB scenario predicts tachyonic sleptons so that the minimal theory must be extended.
There are several ways to tackle the tachyonic slepton problem \cite{tachyonslepton}.
A very elegant solution is the deflected AMSB \cite{deflect} scenario, in which additional
messenger sectors are introduced to deflect the Renormalization Group Equation (RGE)
trajectory and give new contributions to soft SUSY breaking terms\cite{okada,Hsieh:2006ig,Luty:2001zv,Nelson:2002sa,Everett:2008qy}.
On the other hand, a relatively large number of messenger species are needed to
give positive slepton masses with small negative deflection parameters.
It is known that too many messenger fields may lead to strong gauge couplings below GUT scale
or Landau pole below Planck scale. So it is preferred to introduce
less messenger species to deflect the RGE trajectory and at the same time
give positive slepton masses. In our previous work \cite{Fei:1508.01299},
we proposed to solve this problem by introducing
general messenger-matter interactions in the deflected AMSB which has advantages in several aspects. 

Note that in order to preserve gauge coupling unification, the messenger species
are generally fitted into complete representations of the GUT group.
However, sometimes it is economic and well motivated to introduce incomplete
representations of GUT group, such as the $SU(3)_c$ and $SU(2)_L$ adjoint messengers in GMSB\cite{han,ilia,tianjun}.
The introduction of incomplete representations of messengers, which seems
to spoil successful gauge coupling unification, can be natural in AMSB.
This is due to the $'decoupling~theorem'$ in ordinary anomaly mediation scenario
which states that the simple messenger threshold (by pure mass term) will not deflect the AMSB trajectory.
By assigning different origin for messenger thresholds
(determined by moduli VEV or pure mass term), even a complete GUT group representation
at high energy may seem as $'incomplete'$ in AMSB at low energy.
Therefore, the messengers in incomplete GUT representations should also be considered in the
study of deflected AMSB.

In this work, we propose to study a generalized deflected AMSB scenario involving messenger-matter interactions with incomplete GUT multiplets.  
As noted before, the introduction of incomplete GUT mulitiplets in anomaly-type mediation scenarios can be advantageous in various aspects. 
 Besides, virtues of ordinary deflected AMSB are kept while the undesirable Landau-pole type problems can be evaded. 
Such scenarios can be preferable in solving the muon $g-2$ anomaly.
It is known that a SUSY spectrum with heavy colored
sparticles and light non-colored sparticles is needed in order to solve the muon $g-2$ anomaly  and at the same time
be compatible with the LHC data.
We try to realize such a spectrum in the deflected AMSB scenario with general
messenger-matter interactions, where the messengers can form complete or incomplete
GUT representations.
In our scenario, the slepton sector can receive
additional contributions from both the messenger-matter interactions and ordinary
deflected anomaly mediation to avoid tachyonic slepton masses, while the colored sparticles
can be heavy to evade various collider constraints.

This paper is organized as follows. In Sec 2, we study the
soft parameters in the deflected AMSB scenarios with different messenger-matter interactions.
The explanation of the muon $g-2$ in our scenarios and the relevant numerical results
are presented in Sec 3.  Sec 4 contains our conclusions.

\section{General matter-messenger interactions in deflected AMSB}
It is well known that the ordinary AMSB is bothered with the tachyonic slepton problem.
Deflected AMSB scenario, which can change the RGE trajectory
below the messenger thresholds, can elegantly solve such a problem. However, possible strong couplings
at the GUT scale or the Landau pole problem may arise with a small negative deflection parameter.
Positively deflected AMSB, which may need specific forms of moduli superpotential \cite{okada} or
strong couplings \cite{Fei:1505.02785}, could be favored in certain circumstances. However, our previous study
indicated that the Landau pole problem may still persist with a small positive deflection parameter
in order to solve the $g_\mu-2$ anomaly.

In \cite{Fei:1508.01299}, we proposed to introduce general messenger-matter interactions in the messenger sector
which can have several advantages. In this work, the scenarios with complete or incomplete GUT representation
messengers accompanied by messenger-matter interactions will be studied. Note that, the introduction of both adjoint
messengers in {\bf 3} and {\bf  8} representations of $SU(2)_L$ and $SU(3)_c$, respectively, will not spoil the gauge
coupling unification\cite{han}.

Besides, even if the low energy messenger sector seems to spoil the gauge coupling unification, the UV theory
can still be consistent with the GUT requirement. As noted previously, the decoupling theorem in anomaly mediation
ensures that the vector-like thresholds with pure mass terms $M_T>M_{mess}$ will not affect the AMSB trajectory upon
messenger scales. So each low energy (deflected) AMSB theory with incomplete GUT multiplet messengers below
messenger scale $M_{mess}$ could be UV completed to a high energy theory with completed GUT multiplets  at certain
scale upon $M_{mess}$. Incomplete GUT multiplet messengers can also origin from orbifold GUT models by proper
boundary conditions.

The formulas in deflected AMSB with messenger-matter interactions can be obtained from the wavefunction renormalization approach\cite{giudice}
with superfield wavefunction
\beqa
&&{\cal Z}(\tl{\mu};\tl{X},\tl{X}^\da)=Z(\mu;X,X^\da)+\[\theta^2 F \f{\pa}{\pa X}+\bar{\theta}^2 F^\da \f{\pa}{\pa X^\da}+(\theta^2\tl{F}+\bar{\theta}^2 \tl{F}^\da)\f{\pa}{\pa \mu}\]Z(\mu;X,X^\da)\,\nn\\
&&~~~~~~~~~~~~+\theta^2\bar{\theta}^2\( F^\da F \f{\pa^2}{\pa X\pa X^\da}+ F^\da \tl{F} \f{\pa^2}{\pa X^\da\pa \mu}+ \tl{F}^\da {F} \f{\pa^2}{\pa X \pa \mu}+ \tl{F}^\da\tl{F}\f{\pa^2}{\pa \mu^2}\)Z(\mu;X,X^\da).\nn
\eeqa
After canonically normalize the field
\beqa
Q^\pr\equiv Z^{1/2}\[1+\theta^2 \(\f{F}{M} \f{\pa}{\pa \ln X }+\f{\tl{F}}{\mu}\f{\pa}{\pa \ln \mu}\)\ln Z(\mu;X,X^\da)\],
\eeqa
 we can obtain the sfermion masses for the most general forms of deflected AMSB
\beqa
\label{general}
m^2&=&-\[\(\f{F}{M}\)^2\f{\pa^2}{\pa \ln X^\da \pa \ln X}+\(\f{\tl{F}}{\mu}\)^2\f{\pa^2}{\pa \ln\mu^2 }
\right.\nonumber\\ && \left. ~~~~~~~~
+\f{F \tl{F}}{M \mu}\(\f{\pa^2}{\pa \ln X^\da \pa \ln \mu}+\f{\pa^2}{\pa \ln X \pa \ln \mu }\)\]\ln Z(\mu;X,X^\da)~.
\eeqa
From the canonicalized normalized superpotential
\beqa
{\cal L}&=&\int d^2\theta\sum\limits_{i=a,b,c}\[1-\theta^2 \(F \f{\pa}{\pa X }+\tl{F}\f{\pa}{\pa \mu}\)\ln Z(\mu;X,X^\da)\]
\(Z^{-1/2}Q^\pr_i\)\f{\pa W\[(Z^{-1/2} Q^\pr_i)\]}{\pa (Z^{-1/2} Q^\pr_i)} ~, \nn\eeqa
we can obtain the trilinear soft terms
\beqa
\f{A_{abc}}{y_{abc}}&=&\sum\limits_{i=a,b,c}\(\f{F}{M} \f{\pa}{\pa X }+\f{\tl{F}}{\mu}\f{\pa}{\pa \mu}\)\ln Z_i(\mu;X,X^\da).
\eeqa
In our scenario, we have the following replacement
\beqa
\f{F}{M}\ra d F_\phi ~,\f{\tl{F}}{\mu}\ra -F_\phi/2.
\eeqa
Details on general messenger-matter interactions in deflected AMSB can be found in our previous
work \cite{Fei:1508.01299}.

\subsection{Two scenarios with messenger-matter interactions}

\bit
\item Scenario A: deflected AMSB with complete SU(5) GUT representations messengers.

We introduce the following $'N'$ family of new messengers which are fitted into ${\bf 5}$ and ${\bf \overline{5}}$
representation of SU(5) GUT group to deflect the AMSB trajectory
\beqas
 &&\overline{Q}^{I}_\phi(1,2)_{1/2}, ~~\tl{Q}^{I}_\phi(1,\bar{2})_{-1/2},
 ~~\overline{T}^{I}_\phi(\bar{3},1)_{1/3}, ~~{T}^{I}_\phi(3,1)_{-1/3},~~~(I=1,\cdots,N)
 \eeqas
 We introduce the following superpotential that involves messenger-MSSM-MSSM interaction,
typically the slepton-slepton-messenger interaction:
 \beqa
 W&=&\sum\limits_{I}\(\la_A S \overline{Q}^I_\phi \tl{Q}^I_\phi + \la_B S \overline{T}^I_\phi T^I_\phi\)+ \la_X S \overline{Q}^A_\phi \tl{H}_d+ W(S)\nn\\
 &&+\sum\limits_{i,j} \[ \tl{y}^E_{ij} L_{L,i}\tl{Q}^A_\phi  E_{L,j}^c+ \tl{y}^D_{ij} Q_{L,i}\tl{H}_d D_{L,j}^c+y^U_{ij} Q_{L,i}{H}_u U_{L,j}^c\],
 \eeqa
 with certain form of superpotential $W(S)$ for pseduo-moduli field $S$ to determine the deflection parameter $d$.
From the form of the interaction, we can see that the slepton soft SUSY breaking parameters will be different
from the ordinary deflected AMSB results.

\item Scenario B: deflected AMSB with incomplete SU(5) GUT representations messengers.

Motivated by the GMSB with adjoint messenger scenario, we introduce the following incomplete SU(5) GUT
representation messengers to deflect the AMSB trajectory
\beqas
  &&\Sigma^I_O(8,1)_0,~~\sigma^I_T(1,3),~~Z^{J}(1,1)_{1},~~\bar{Z}^{J}(1,1)_{-1},~~I,J=(1,\cdots,M)~;\\
 &&\overline{Q}^{A}_\phi(1,2)_{1/2}, \tl{Q}^{A}_\phi(1,\bar{2})_{-1/2},~\overline{T}^{A}_\phi(\bar{3},1)_{1/3}, {T}^{A}_\phi(3,1)_{-1/3}.
\eeqas
We note that additional singlet messengers $Z^I$ with non-trivial $U(1)_Y$ quantum number can be introduced to
deflected the $\tl{E}_L^c$ slepton RGE trajectory.
As in the previous scenario, the superpotential also involves messenger-MSSM-MSSM interaction, typically the
slepton-slepton-messenger interaction:
 \beqa
 W&=&\la_A S \overline{Q}^A_\phi \tl{Q}^A_\phi +\la_B S \overline{T}^A_\phi T^A_\phi+ \sum\limits_{I}\[\la_O S Tr(\Sigma_O^I \Sigma^I_O)+ \la_T S Tr(\Sigma^I_T \Sigma^I_T) \right. \nn \\
&& \left. +\la_Z S \overline{Z}^I Z^I\]
   +\la_X S \overline{Q}^A_\phi \tl{H}_d+\sum\limits_{i,j} \[ \tl{y}^E_{ij} L_{L,i}\tl{Q}^A_\phi  E_{L,j}^c
+ \tl{y}^D_{ij} Q_{L,i}\tl{H}_d D_{L,j}^c\right. \nn \\
&& \left.
+y^U_{ij} Q_{L,i}{H}_u U_{L,j}^c\]+ W(S)
 \eeqa
\eit
 We can see that there will be mixing between the messenger $\tl{Q}^A_\phi$ and $\tl{H}_d$
(as well as  $Q^B_\phi$ and $H_u$). We will define the new states
 \beqa
 {Q}^A_\phi&\equiv&\f{\la_A \tl{Q}^A_\phi+ \la_X \tl{H}_d}{\sqrt{\la_A^2+\la_X^2}}~,~~~~~
 H_d\equiv\f{-\la_X \tl{Q}^A_\phi+ \la_A \tl{H}_d}{\sqrt{\la_A^2+\la_X^2}}~~.
  \eeqa
After the substitution of the new states, the superpotential changes to
\beqa
W&=&\sqrt{\la_A^2+\la_X^2} S \overline{Q}^A_\phi {Q}^A_\phi+ \sum\limits_{i,j}y^U_{ij} Q_{L,i} {H}_u U_{L,j}^c+\sum\limits_{F_H=T^I_\phi,Q_\phi^I,\cdots}\la_{F_H} S \bar{F}_H F_H+ W(S)
 ~\nn\\
 &+&\sum\limits_{i,j}
\tl{y}^E_{ij} L_{L,i} \f{\la_A Q_\phi^A-\la_X H_d}{\sqrt{\la_A^2+\la_X^2}}  E_{L,j}^c+  \tl{y}^D_{ij} Q_{L,i} \f{\la_X Q_\phi^A+\la_A H_d}{\sqrt{\la_A^2+\la_X^2}} D_{L,j}^c~,
\eeqa
We have the following relation
\beqa
\tl{y}^E_{ij}\f{\la_X}{\sqrt{\la_A^2+\la_X^2}}=y^E_{ij}~,~~~-\tl{y}^D_{ij}\f{\la_A}{\sqrt{\la_A^2+\la_X^2}}=y^D_{ij}~.
\eeqa
We define
\beqa
\tl{y}^E_{ij}\f{\la_A}{\sqrt{\la_A^2+\la_X^2}}\equiv-\la^E_{ij}~,~~~-\tl{y}^D_{ij}\f{\la_X}{\sqrt{\la_A^2+\la_X^2}}\equiv\la^D_{ij}~,~~~ \sqrt{\la_A^2+\la_X^2}\equiv\la_S~.
\eeqa
So the superpotential can be rewritten as
\beqa
W&=&\la_S S \overline{Q}^A_\phi {Q}^A_\phi+ \sum\limits_{i,j}y^U_{ij} Q_{L,i} {H}_u U_{L,j}^c+\sum\limits_{F_H=Z^I,Q^I,\cdots}\la_{F_H} S \bar{F}_H F_H+ W(S) \nn\\
 &-&\sum\limits_{i,j}
\[\la^E_{ij} L_{L,i} Q_\phi^A E_{L,j}^c+y^E_{ij} L_{L,i} H_d E_{L,j}^c+\la^D_{ij} Q_{L,i} Q_\phi^A D_{L,j}^c+y^D_{ij} Q_{L,i} H_d D_{L,j}^c
\]\eeqa
For simplicity, we chose $\la_{ij}^E=\la_E \delta_{ij}, \la^D_{ij}=\la_D \delta_{ij}$ to be diagonal.
Below the messenger threshold determined by the VEV of pseudo-moduli $S$, we can integrate out the
heavy fields $F_H,Q_\phi^A$ and obtain the low energy MSSM.

\subsection{The soft SUSY spectrum in two scenarios}
From the superpotential, the soft SUSY breaking parameters can be calculated. In the calculation, the wavefunction
renormalizatin approach \cite{wavefunction:hep-ph/9706540} is used in which messenger threshold $M_{mess}^2$ is
replaced by spurious chiral fields $X$ with $M_{mess}^2=X^\da X $.
The most general type of expressions in AMSB can be found in our previous work \cite{Fei:1508.01299}.

We can calculate the change of the gauge beta-function
\beqa
&&\Delta \beta_{g_i}=\f{1}{16\pi^2}g_i^3\Delta b_{g_i}~,
\eeqa
with
\beqa
&&\Delta (b_3,b_2,b_1)=(~N,~N,~N),~~~~~~~~~~~~~~~
\eeqa
for Scenario A.
For Scenario B we consider two cases.
One is
\beqa
&&\Delta (b_3,b_2,b_1)=(~3M+1,~2M+1,~1)~~~~~~~{\rm Scenario~B1}~
\eeqa
in which $'I=M,J=0'$ is adopted to guarantee apparently gauge coupling unification.
The other is
\beqa
&&\Delta (b_3,b_2,b_1)=(~3M+1,~2M+1,~\f{6M}{5}+1)~~~~~{\rm Scenario~B2}~
\eeqa
with $'I=J=M'$ in which apparently the gauge coupling unification is spoiled.
However, as we discussed previously, successful GUT may still be possible if certain additional
incomplete messengers upon $'X'$ threshold determined by pure mass terms are introduced in the UV completed theory.

From the general expressions in Eq.(\ref{general}), we can see that there are three types of contributions
to the soft SUSY breaking parameters:
\bit
\item The interference contribution part given by
\beqa
\delta^I&=&\f{\pa^2}{\pa\ln \mu\pa\ln X}\ln Z^D_{ab}~\nn\\
 &=&\f{\pa^2}{\pa\ln \mu\pa\ln X}Z^D-\f{\pa}{\pa\ln\mu}Z^D\f{\pa}{\pa \ln X}Z^D~\nn\\
&=&(\f{\Delta G^D_a}{2}\f{\pa }{\pa Z_a^D}+\f{\Delta\beta_{g_r}}{2}\f{\pa}{\pa g_r}){G^-}-G_a^D\f{\Delta G_a}{2}~.
\eeqa
In our convention, the anomalous dimensions are expressed in the holomorphic basis \cite{shih,chacko}
\beqa
G^i\equiv \f{d Z_{ij}}{d\ln\mu}\equiv-\f{1}{8\pi^2}\(\f{1}{2}d_{kl}^i\la^*_{ikl}\la_{jmn}Z_{km}^{-1*}Z_{ln}^{-1*}-2c_r^iZ_{ij}g_r^2\),
\eeqa
We define $(\Delta G\equiv G^+ - G^-)$, the discontinuity across the integrated heavy field threshold
with $G^+(G^-)$ denoting the value upon (below) such threshold, respectively.

The discontinuities of the relevant couplings are given as
\beqa
\Delta G_{y_t}&=&-\f{1}{8\pi^2}\(\la_D^2\)~,\\
\Delta G_{y_b}&=&-\f{1}{8\pi^2}\(3\la_D^2\)~,\\
\Delta G_{y_\tau}&=&-\f{1}{8\pi^2}\(3\la_E^2\)~
\eeqa
We take into account the terms involving $y_t,y_b,y_\tau, g_i,\lambda$, and the subleading terms
are neglected in the calculation.
The new interference contributions from the messenger-matter interactions are given as
\beqa
2\delta^I_{Q_{L,i}}&=&\delta_{i,3}\f{d F_\phi^2}{8\pi^2}\[y_t^2\Delta G_{y_t}+y_b^2\Delta G_{y_b}\]~,\\
2\delta^I_{U_{L,i}^c}&=&\delta_{i,3}\f{d F_\phi^2}{8\pi^2}\[2y_t^2\Delta G_{y_t}\]~,\\
2\delta^I_{D_{L,i}^c}&=&\delta_{i,3} \f{d F_\phi^2}{8\pi^2}\[2y_b^2\Delta G_{y_b}\]~,
\eeqa
\beqa
2\delta^I_{L_{L,i}}&=&\delta_{i,3} \f{d F_\phi^2}{8\pi^2}\[y_\tau^2\Delta G_{y_\tau} \]~,\\
2\delta^I_{E_{L,i}^c}&=&\delta_{i,3} \f{d F_\phi^2}{8\pi^2}\[2y_\tau^2\Delta G_{y_\tau}\]~,\\
2\delta^I_{H_u}&=&\f{d F_\phi^2}{8\pi^2}\[3y_t^2\Delta G_{y_t}\]~,\\
2\delta^I_{H_d}&=& \f{d F_\phi^2}{8\pi^2}\[3y_b^2\Delta G_{y_b}+ y_\tau^2\Delta G_{y_\tau}\]~,
\eeqa
with $\delta_{i,j}$ the Kronecker delta. Terms involving the gauge parts are absorbed in the deflected AMSB contributions involving $G_i$.

\item  The pure gauge mediation part given by
\beqa
\delta^G=\f{\pa^2}{\pa\ln X\ln X^\da}\ln Z^D=\f{\pa^2}{\pa \ln X \ln X^\da} Z^D- \f{\pa  Z^D}{\pa \ln X}\f{\pa  Z^D}{\pa \ln X^\da}~.
\eeqa
Note that
\beqa
\Delta G_{Q_iQ_i}&=&-\f{1}{8\pi^2}\[\la_D^2\]~,\\
\Delta G_{D_iD_i}&=&-\f{1}{8\pi^2}\[2\la_D^2\]~,\\
\Delta G_{E_iE_i}&=&-\f{1}{8\pi^2} 2\la_E^2~,\\
\Delta G_{L_iL_i}&=&-\f{1}{8\pi^2} \la_E^2~,
\eeqa
and
\beqa
G^+_{L_iL_i}&=&-\f{1}{8\pi^2}\[\la_E^2+y_\tau^2\delta_{i,3}\]~,\\
G^+_{E_iE_i}&=&-\f{1}{8\pi^2}2\[\la_E^2+y_\tau^2\delta_{i,3}\]~,\\
G^+_{Q_iQ_i}&=&-\f{1}{8\pi^2}\[\la_D^2+(y_b^2+y_t^2)\delta_{i,3}\]~,\\
G^+_{D_iD_i}&=&-\f{1}{8\pi^2}2\[\la_D^2+y_b^2\delta_{i,3}\]~,\\
G^+_{Q_\phi Q_\phi}&=&-\f{1}{8\pi^2}\[\la_E^2+3\la_D^2+\la_S^2\]~,
\eeqa
and also the anomalous dimension above the messenger threshold
\beqa
G_{\la^D_{ii}}^+&=&G_{Q_iQ_i}^++G_{D_iD_i}^++G^+_{Q_\phi Q_\phi}\nn\\
&=&-\f{1}{8\pi^2}\[6\la_D^2+\la_E^2+\la_S^2+(3 y_b^2+y_t^2)\delta_{i,3}-\f{16}{3}g_3^2-3g_2^2-\f{7}{15}g_1^2\]~,\\
G_{\la^E_{ii}}^+&=&G_{L_iL_i}^++G_{E_iE_i}^++G^+_{Q_\phi Q_\phi}\nn\\
&=&-\f{1}{8\pi^2}\[4\la_E^2+3 \la_D^2+\la_S^2+3y_\tau^2\delta_{i,3}-3g_2^2-\f{9}{5}g_1^2\]~,
\eeqa
so we have
\beqa
4\delta^G_{Q_i }&=&\f{d^2 F_\phi^2}{8\pi^2}\[\la_D^2G_{\la^D_{ii}}^+\]-\f{d^2 F_\phi^2}{8\pi^2}\delta_{i,3}\[y_t^2\Delta G_{y_t}+y_b^2\Delta G_{y_b}\]~,\\
4\delta^G_{D_i }&=&\f{d^2 F_\phi^2}{8\pi^2}\[2\la_D^2G_{\la^D_{ii}}^+\]-\f{d^2 F_\phi^2}{8\pi^2}\delta_{i,3}\[2y_b^2\Delta G_{y_b}\]~,\\
4\delta^G_{U_i }&=&-\f{d^2 F_\phi^2}{8\pi^2}\delta_{i,3}\[2y_t^2\Delta G_{y_t}\]~,\\
4\delta^G_{L_i }&=&\f{d^2 F_\phi^2}{8\pi^2}\[\la_E^2G_{\la^E_{ii}}^+\]-\f{d^2 F_\phi^2}{8\pi^2}\delta_{i,3}\[y_\tau^2\Delta G_{y_\tau}\]~,\\
4\delta^G_{E_i }&=&\f{d^2 F_\phi^2}{8\pi^2}\[2\la_E^2G_{\la^E_{ii}}^+\]-\f{d^2 F_\phi^2}{8\pi^2}\delta_{i,3}\[2y_\tau^2\Delta G_{y_\tau}\]~,\\
4\delta^G_{H_u }&=&-\f{d^2 F_\phi^2}{8\pi^2}\[3y_t^2\Delta G_{y_t}\],\\
4\delta^G_{H_d }&=&-\f{d^2 F_\phi^2}{8\pi^2}\[3y_b^2\Delta G_{y_b}+y_\tau^2\Delta G_{y_\tau}\]
\eeqa

\item The pure deflected AMSB contributions without messenger-matter interactions given by
\beqa
\delta_A&=&\f{d^2 Z^D}{dt^2}-\(\f{dZ^D}{dt}\)^2~ \nn\\
&=&\(G_a^D\f{\pa}{\pa Z_a}+\beta_g^i\f{\pa}{\pa g_i}\)G^D-(G_D^2)~,
\eeqa
The expressions are given by
\beqa
\delta_{\tl{Q}_{L,i}}^A&=&\f{F_\phi^2}{16\pi^2}\[\f{8}{3} G_3 \al^2_3+\f{3}{2}G_2\al^2_2+\f{1}{30}G_1\al^2_1\]
 \nn\\
&& +\delta_{3,i}\f{F_\phi^2}{(16\pi^2)^2}y_t^2(6y_t^2+y_b^2-\f{16}{3}g_3^2-3g_2^2-\f{13}{15}g_1^2) \nn\\
&& +\delta_{3,i}\f{F_\phi^2}{(16\pi^2)^2}y_b^2(y_t^2+6y_b^2+y_\tau^2-\f{16}{3}g_3^2-3g_2^2-\f{7}{15}g_1^2)~,\\
\delta_{\tl{U}^c_{L,i}}^A&=&\f{F_\phi^2}{16\pi^2}\[\f{8}{3} G_3 \al^2_3+\f{8}{15}G_1\al^2_1\]
+\delta_{3,i} \f{F_\phi^2}{(16\pi^2)^2}2y_t^2(6y_t^2+y_b^2-\f{16}{3}g_3^2-3g_2^2-\f{13}{15}g_1^2)~,\nn
\eeqa\beqa
{\delta_{\tl{D}^c_{L,i}}^A}&=&\f{F_\phi^2}{16\pi^2}\[\f{8}{3} G_3 \al^2_3+\f{2}{15}G_1\al^2_1\]
\nn\\
&& +\delta_{3,i} \f{F_\phi^2}{(16\pi^2)^2}2y_b^2(y_t^2+6y_b^2+y_\tau^2-\f{16}{3}g_3^2-3g_2^2-\f{7}{15}g_1^2)~,\\
\delta^A_{\tl{H}_u} &=&\f{F_\phi^2}{16\pi^2}\[\f{3}{2}G_2\al^2_2+\f{3}{10}G_1\al^2_1\]\nn\\
&& +\f{F_\phi^2}{(16\pi^2)^2}3y_t^2(6y_t^2+y_b^2-\f{16}{3}g_3^2-3g_2^2-\f{13}{15}g_1^2),\\
\delta^A_{\tl{H}_d} &=&\f{F_\phi^2}{16\pi^2}\[\f{3}{2}G_2\al^2_2+\f{3}{10}G_1\al^2_1\]\nn \\
&& +\f{F_\phi^2}{(16\pi^2)^2}3y_b^2(y_t^2+6y_b^2+y_\tau^2-\f{16}{3}g_3^2-3g_2^2-\f{7}{15}g_1^2)\nn\\
&& +\f{F_\phi^2}{(16\pi^2)^2} y_\tau^2\(4y_\tau^2+3y_b^2-3g_2^2-\f{9}{5}g_1^2\)~,\\
\delta^A_{\tl{L}_{L,i}} &=&\f{F_\phi^2}{16\pi^2}\[\f{3}{2}G_2\al^2_2+\f{3}{10}G_1\al^2_1\]\nn\\
&& +\delta_{3,i}\f{F_\phi^2}{(16\pi^2)^2} y_\tau^2\(4y_\tau^2+3y_b^2-3g_2^2-\f{9}{5}g_1^2\), \\
\delta^A_{\tl{E}_{L,i}^c} &=&\f{F_\phi^2}{16\pi^2}\[\f{6}{5}G_1\al^2_1\]+\delta_{3,i}\f{F_\phi^2}{(16\pi^2)^2}2y_\tau^2\(4y_\tau^2+3y_b^2-3g_2^2-\f{9}{5}g_1^2\)~,
\eeqa
with
\beqa
G_i=(\Delta b_i) d^2+2(\Delta b_i) d-b_i~,\\
(b_1,b_2,b_3)=(\f{33}{5},1,-3)~~.
\eeqa
\eit

So we obtain the final results of soft SUSY breaking parameters for sfermions
\beqa
{m^2_i}&=&-\delta_i^I+\delta_i^G+\delta_i^A~,
\eeqa
with $'d'$ being the deflection parameter.

The trilinear coupling will also receive new contributions which are given by
\beqa
A_t&=&\f{F_\phi}{16\pi^2}\[6y_t^2+y_b^2-(\la_D^2) d-\f{16}{3}g_3^2-3g_2^2-\f{13}{15}g_1^2\]~,\\
A_b&=&\f{F_\phi}{16\pi^2}\[y_t^2+6y_b^2+y_\tau^2-(3\la_D^2) d-\f{16}{3}g_3^2-3g_2^2-\f{7}{15}g_1^2\]~,\\
A_\tau&=&\f{F_\phi}{16\pi^2}\[4y_\tau^2+3y_b^2-(3\la_E^2) d -3g_2^2-\f{9}{5}g_1^2\]~,\\
A_{U;1,2}&=&\f{F_\phi}{16\pi^2}\[-(\la_D^2) d-\f{16}{3}g_3^2-3g_2^2-\f{13}{15}g_1^2\]~,\eeqa\beqa
A_{D;1,2}&=&\f{F_\phi}{16\pi^2}\[-(3\la_D^2) d-\f{16}{3}g_3^2-3g_2^2-\f{7}{15}g_1^2\]~,\\
A_{E;1,2}&=&\f{F_\phi}{16\pi^2}\[-(3\la_E^2) d -3g_2^2-\f{9}{5}g_1^2\]~.
\eeqa

The gaugino masses are determined by
\beqa
{m_{\la_i}}&=&{g^2}\f{F_\phi}{2}\(\f{\pa}{\pa\ln\mu}-d\f{\pa}{\pa \ln |X|}\)\f{1}{g^2(\mu,X)} \nn\\
&=&{g^2}\f{F_\phi}{2}\(2\f{1}{16\pi^2}b_i-2d\f{1}{16\pi^2}\Delta b_i\) \nn\\
&=&{g^2}\f{F_\phi}{16\pi^2}\( b_i-d \Delta b_i\)~.
\eeqa
So we have
\beqa
\label{gauginomass}
m_{\la_i}=\f{F_\phi}{4\pi}\al_i\(b_i-d \Delta b_i\).
\eeqa
Therefore, the gaugino masses at the messenger scale are given as
\beqa
 M_3&=& \f{F_\phi}{4\pi}\al_3 \[-3-d(\Delta b_3)\]~, \\
 M_2&=& \f{F_\phi}{4\pi}\al_2\[1-d(\Delta b_2)\]~, \\
 M_1&=& \f{F_\phi}{4\pi}\al_1 \[6.6-d (\Delta b_1) \]~.
\eeqa

It is well known in AMSB that naively adding a supersymmetric $\mu$ term to the Lagrangian will lead
to unrealistic large $B\mu= \mu F_\phi$.
So the generations of $\mu$ and $B_\mu$ in AMSB may have a different origin and are model dependent.
In fact, there are already many proposals to generate realistic $\mu$ and $B\mu$, for example, by
promoting to NMSSM \cite{cao} or introducing a new singlet \cite{1008.2024}. We will treat them as
free parameters in this scenario.

\section{Solving the muon g-2 anomaly in our scenario}

The E821 experimental result of the muon anomalous magnetic moment at the Brookhaven AGS \cite{ex:g-2}
\beqa
a_\mu^{\rm expt} =116592089(63)\times 10^{-11}~,
\eeqa
is larger than the SM prediction\cite{sm:g-2}
\beqa
a^{\rm SM}_\mu =116591834(49)\times 10^{-11}~.
\eeqa
 The deviation is about $3.2 \sigma$
 \beqa
\Delta a_\mu({\rm expt - SM}) = (255\pm 80)\times 10^{-11}.
\eeqa
SUSY can yield sizable contributions to the muon $g-2$ which dominantly
come from the chargino-sneutrino and the neutralino-smuon loop diagrams.
The muon $g-2$ anomaly, which is order $10^{-9}$,  can be explained for $m_{\rm SUSY} = {\cal O}(100)$ GeV
and $\tan\beta = {\cal O}(10)$. In our scenario, slepton masses as well as $M_1,M_2$ can be relatively light.
On the other hand, the colored sparticles can be heavy to evade possible constraints from the LHC, the
SUSY flavor and CP problems. Some recent discussions can be seen in \cite{muon:g-2}.

The soft terms are characterized by the following free parameters at the messenger scale
  \beqa
  d, M_{mess}, F_\phi, \tan\beta, \la^D,~\la^E,~\la_S,~\la_{F_H}
  \eeqa
All the inputs should be seen as the boundary conditions at the messenger scale,
which after RGE running to the EW scale, could give the low energy spectrum.
About these parameters, we have the following comments:
\bit
\item  The value of $F_\phi$ is chosen to lie in the range $1{\rm TeV}<F_\phi<500 {\rm TeV}$.
We know that the value of $F_\phi$ determines the whole spectrum. On the one hand, $F_\phi$ cannot be very
low due to the constraints from the gaugino masses. A very heavy $F_\phi$ will spoil
the EWSB requirement and give a Higgs mass heavier than the LHC results.
\item The messenger scale $M_{mess}$ can be chosen to be less than the GUT scale and at the same time
heavier than the sparticle spectrum.   So we choose $ 1 {\rm TeV} \leq M_{mess}\leq 10^{15} {\rm GeV}$.
\item  We choose the deflection parameter in the range $-5\leq d\leq 5$ and $\tan\beta$ in the range
       $2\leq\tan\beta\leq 50$.
\item  The parameters $\la_D,\la_E,\cdots$ can be chosen in the range $0<|\la|<\sqrt{4\pi}$ which ensure
positive contributions to slepton masses regardless of the (sign of) deflection parameter $d$.
This is the advantage of our scenario which needs less messenger species with a given $d$.
\eit
  We also take into account the following collider and dark matter constraints:
\bit
\item[(1)] The mass range for the Higgs boson $123 {\rm GeV}<M_h <127 {\rm GeV}$ from ATLAS and CMS \cite{ATLAS:higgs,CMS:higgs}.
\item[(2)]  The lower bounds on neutralino and charginos masses, including the invisible decay bounds
            for $Z$-boson \cite{EW-precision}.
\item[(3)] The dark matter relic density from the Planck result $\Omega_{DM} = 0.1199\pm 0.0027$ \cite{planck}
           (in combination with the WMAP data \cite{wmap}) and the limits of the LUX-2016\cite{Akerib:2016vxi},the PandaX\cite{PANDAX} spin-independent
           dark matter scattering cross section .
\item[(4)]  Flavor constraints from the rare decays of B-mesons
\bit
\item   Constraints from $Br(B_s \ra \mu^+\mu^-)$\cite{bsmu}
 \beqa
 1.6\times 10^{-9} \leq Br(B_s \ra \mu^+\mu^-)\leq 4.2\tm 10^{-9}~(2\sigma)~,
 \eeqa
 \item  Constraints from $Br(B_S\ra X_s \gamma)$ etc\cite{bsg}
  \beqa
   2.99\tm 10^{-4} \leq Br(B_S\ra X_s \gamma) < 3.87\tm 10^{-4}~(2\sigma)~.
   \eeqa
\eit   
   
\item[(5)] The electroweak precision obsearvables \cite{pdg}, such as
  \beqa
  \delta M_{W}^{\rm exp} \approx \pm 30{\rm MeV}, ~~\delta \sin \theta_{\rm eff}^{\rm exp} \approx \pm 15\times10^{-5}.
   \eeqa
\item[(6)]  Current LHC constraints on sparticle masses \cite{SUSYmass}:
  \bit
    \item  Gluino mass $m_{\tl{g}} \gtrsim 1.5\sim 1.9$ TeV;
    \item  Light stop mass $m_{\tl{t}_1} \gtrsim 0.85$ TeV;
    \item  Light sbottom mass $m_{\tl{b}_1} \gtrsim 0.84$ TeV;
    \item  First two generation squarks $m_{\tl{q}} \gtrsim 1.0 \sim 1.4$ TeV.
    \eit
\eit

\begin{figure}[htb]
\begin{center}
\includegraphics[width=4.0in]{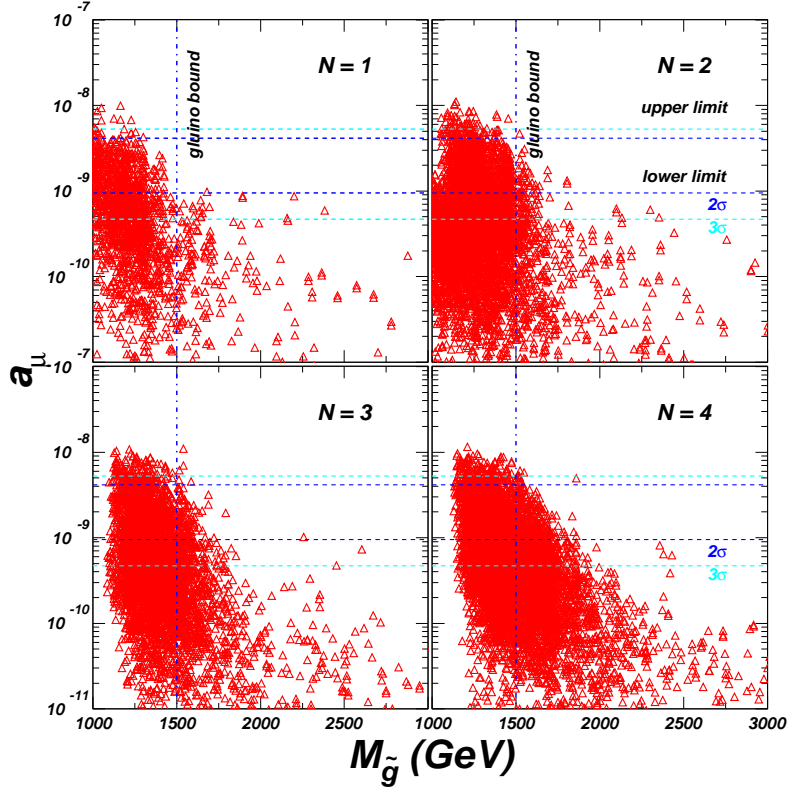}
\end{center}
\vspace{-.5cm}
\caption{The scatter plots of the survived samples showing
the muon $g-2$ versus the gluino mass in Scenario A with complete GUT multiplets.
The blue (cyan) dash line indicate the $2\sigma$ ($3\sigma$) range of the muon $g-2$ data.
A gluino lower bound $m_{\tl{g}}\gtrsim 1.5$ TeV is shown in the figure.}
\label{scA1}
\end{figure}

From the numerical results, we have the following observations:
\bit
\item Scenario A:
Fig.\ref{scA1} shows the scan results of Scenario A  in which the $\Delta a_\mu$ versus $m_{\tilde g}$ plots
with complete GUT multiplets are given. The blue (cyan) dashed line indicate the $2\sigma$ ($3\sigma$) range
of $g_\mu-2$ data.  All survived points satisfy the constraints (1-6) except the bounds from the dark matter
relic density and the gluino mass.
The most stringent constraints come from the LHC bounds on gluino mass, which
excluded a great majority of the survived points that solve the $g_\mu-2$ anomaly at $2\sigma$ level.
As the messenger species number $N$ gets larger, more and more points can survive the gluino mass bound.

\begin{figure}[htb]
\begin{center}
\includegraphics[width=2.9 in]{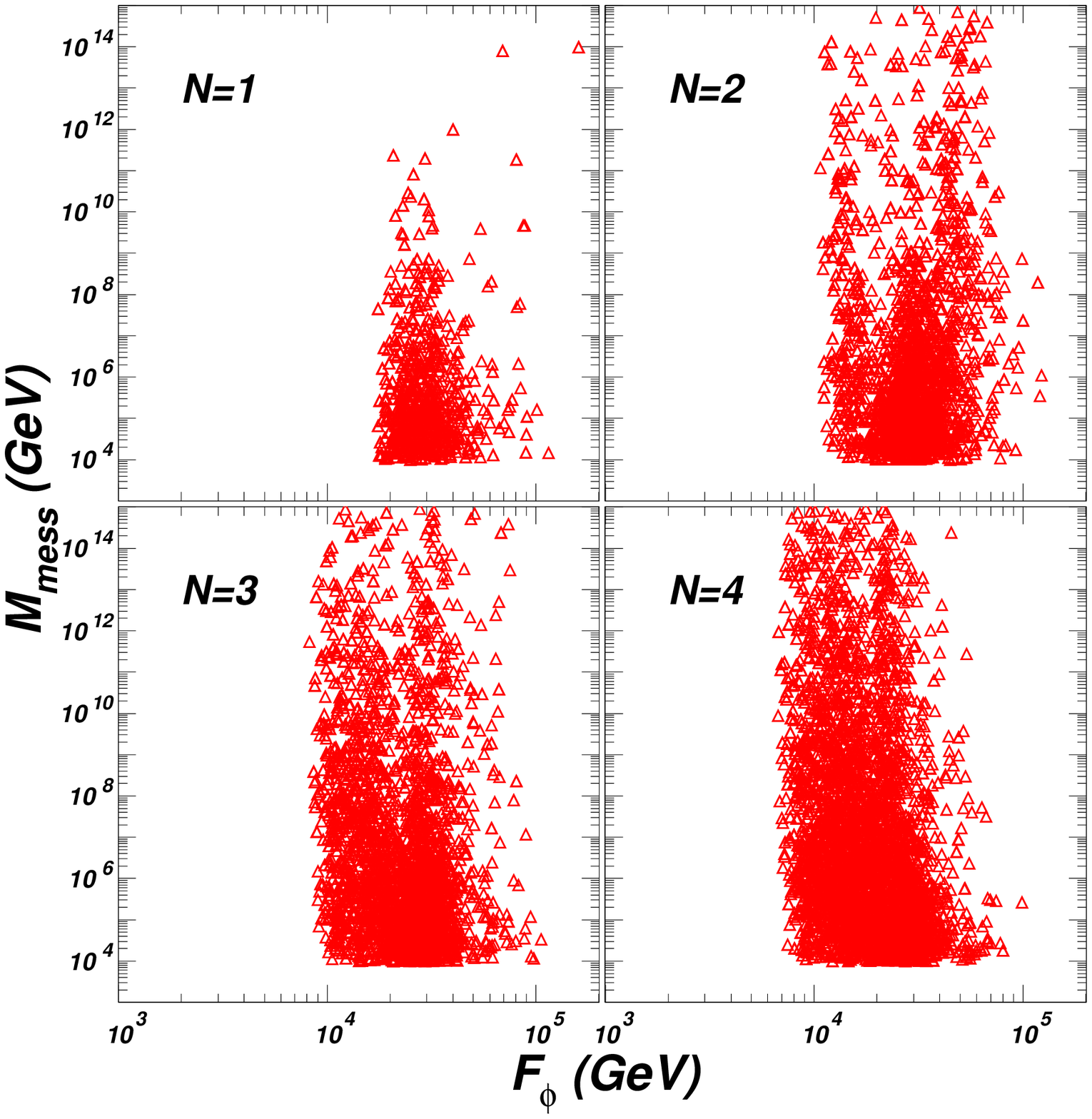}
\includegraphics[width=2.9 in]{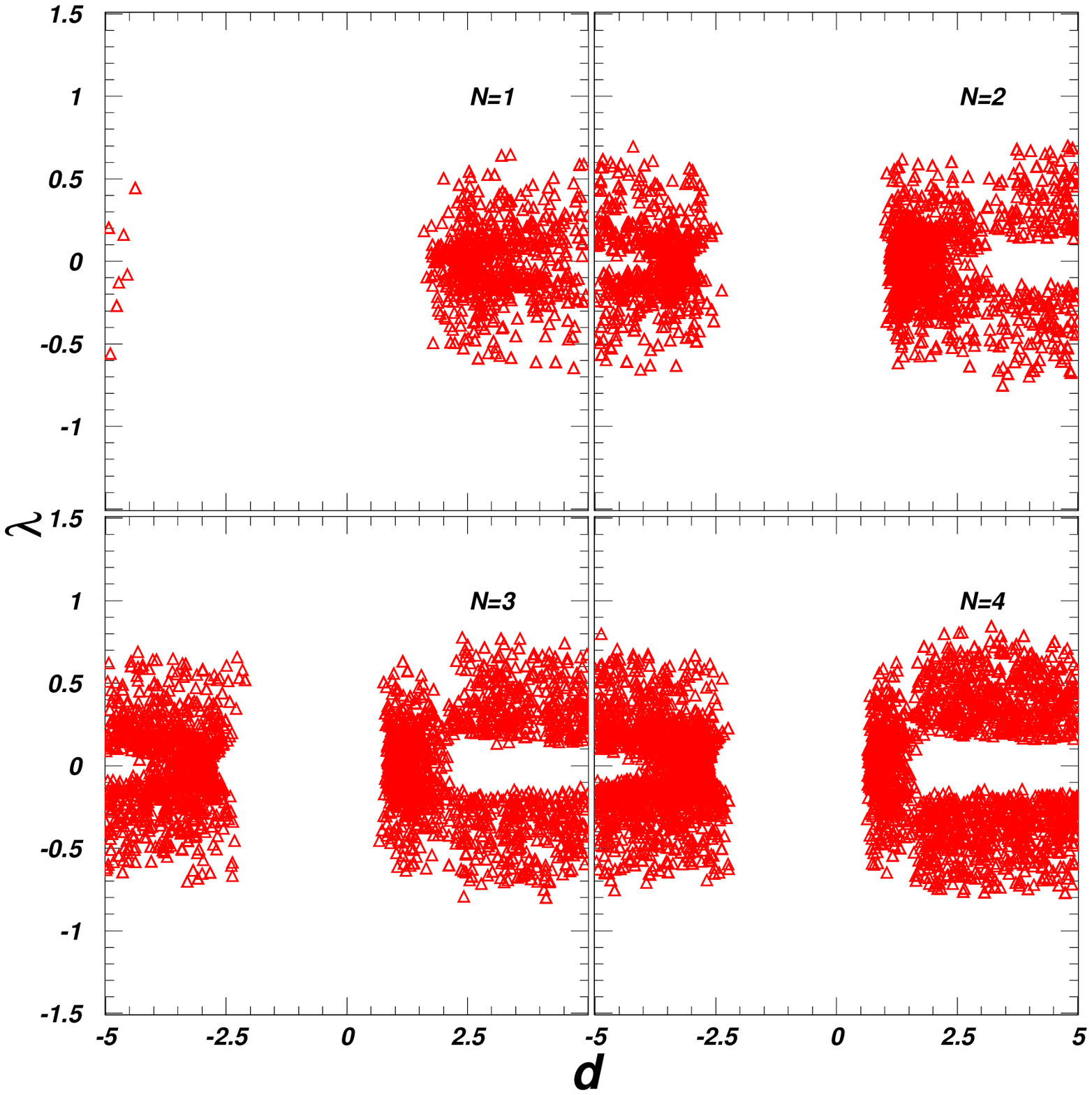}
\end{center}
\vspace{-.8cm}
\caption{Same as Fig.1, but showing the value of $F_\phi$ versus the messenger scale $M_{mess}$ (the left panel)
 and the deflection parameter $d$ versus the messenger-matter couplings $\lambda\equiv \lambda_E$
(the right panel).}
\label{scA2}
\end{figure}

The gluino is upper bounded by about 2.5 TeV (2.0 TeV) if the $g_\mu-2$ anomaly is solved at $3\sigma$ ($2\sigma$)
level. We know that the $g_\mu-2$ anomaly can be solved if the relevant sparticles
$\tl{\mu},\tl{\nu}_\mu,\tl{B},\tl{W}$ are lighter than $600\sim 700$ GeV \cite{muong-2:0208067}
(the region with a smaller $\tan\beta$ needs even lighter sparticles).
In AMSB, the whole low energy spectrum is determined by the value of $F_\phi$.
So, in order to solve the $g_\mu-2$ anomaly, the mass scale of $\tl{\mu},\tl{\nu}_\mu,\tl{B},\tl{W}$
determines the upper bound of $F_\phi$, which, on the other hand, sets a bound on gluino mass.
The allowed range of $F_\phi$ versus the messenger scale $M_{mess}$ in Scenario A is shown in
the left panel of Fig \ref{scA2}. It is obvious from the plots that the scale of $F_\phi$ is indeed upper bounded
to account for the $g_\mu-2$ anomaly.
We should note that the deflection of the RGE trajectory and the messenger-matter interactions can loosen the
bound of $F_\phi$ in comparison with the ordinary AMSB.

The deflection parameter $d$ versus the messenger-matter couplings $\lambda_E\equiv \lambda$ is plotted in the
right panel of Fig.\ref{scA2}.
We see that additional messenger-matter interactions are welcome to explain the $g_\mu-2$ anomaly.
Only a small range of $d$ is allowed without leptonic messenger-matter interactions ($\la_E=0$).
However, the allowed range for $d$ enlarges with non-trivial messenger-matter interactions.

Our numerical results indicate that the majority part of the allowed parameter space can not satisfy the
the upper bound of dark matter relic density.
This result can be understood from the hierarchies among the gauginos at the EW scale.
From Eq.(\ref{gauginomass}), the gaugino mass ratios at the weak scale are given by
 \beqa
 \label{ratio}
 M_1:M_2:M_3\approx \[6.6-d (\Delta b_1) \]:2 \[1-d(\Delta b_2)\]:6\[-3-d(\Delta b_3)\]~.
 \eeqa
 Knowing the range of the deflection parameter $d$, the lightest gaugino can be identified.

 It can be seen in case $N=1$ that the deflection parameter $d$ is lower bounded to $d\gtrsim 1.5$ for a positive $d$
while $d\lesssim-4.5$ for a negative $d$. From Eq.(\ref{ratio}) we can see that for $-4.6<d<2.8$ the lightest gaugino
will be the wino, otherwise the lightest gaugino will be the bino. It is well known that the relic density constraints for bino-like dark matter is very stringent and possible co-annihilation with sleptons or resonance are needed to obtain the correct DM relic density.
So in a majority of the parameter space allowed by $g_\mu-2$ and gluino mass bound,
the LSP will be bino-like and can hardly give the right DM relic density. On the other hand, a small portion of the allowed
parameter space will predict a wino-like LSP which will lead to insufficient dark matter abundance
for a wino mass below 3 TeV unless other DM components (for example, axion) will be present. Heavy wino-like LSP of order 3 TeV will always lead to heavy bino and sleptons which otherwise can not explain the $g_\mu-2$ anomaly. Given the upper bounds on $F_\phi$ from $g_\mu-2$ and
gluino mass, the wino will always be much lighter than 3 TeV. We give in Table 1 the range of $d$, within
which the wino will be lighter than bino for various messenger species $N$. We can see that only a small
portion of parameter space with a positive $d$ can satisfy the dark matter relic density upper bound.
The vast parameter space with a bino-like LSP will be stringently constrained by dark matter relic density
upper bound. We checked that a very small region can satisfy such relic density constraints.
So generalized deflected AMSB scenarios with complete GUT representation of messengers are not favored in solving the $g_\mu-2$ discrepancy.

\begin{table}[h]
\caption{The range of $d$ within which the wino will be lighter than bino for various messenger species $N$
in Scenario A.}
\centering
\begin{tabular}{|c|c|c|c|c|}
\hline
 &~N=1&~N=2&~N=3&~N=4\\
\hline
$d$&$-4.6<d<2.86$ &$-2.3<d<1.43$&$-1.53<d<0.95$&$-1.15<d<0.72$\\
\hline
\end{tabular}
\end{table}

We should note that the constraints from the gluino can be alleviated if we introduce pure colored
messenger particles (without $SU(2)_L$ and $U(1)_Y$ quantum numbers). We can see from the expressions
for the soft SUSY parameters that the value of $\Delta b_3$ can essentially control the gluino mass.
More pure colored messenger particles always mean a heavy gluino for a positive deflection
parameter which, on the other hand, may spoil the gauge coupling unification.
As noted in the previous section, the complete representation messengers may seem
$'incomplete'$ at the low energy $X$ threshold. However, the perturbative gauge coupling
unification may be spoiled with more additional messenger species. We will discuss the detailed
consequence of general messenger sectors versus gauge coupling unification in our subsequent
studies.

\begin{figure}[htb]
\label{scB1}
\begin{center}
\includegraphics[width=5 in]{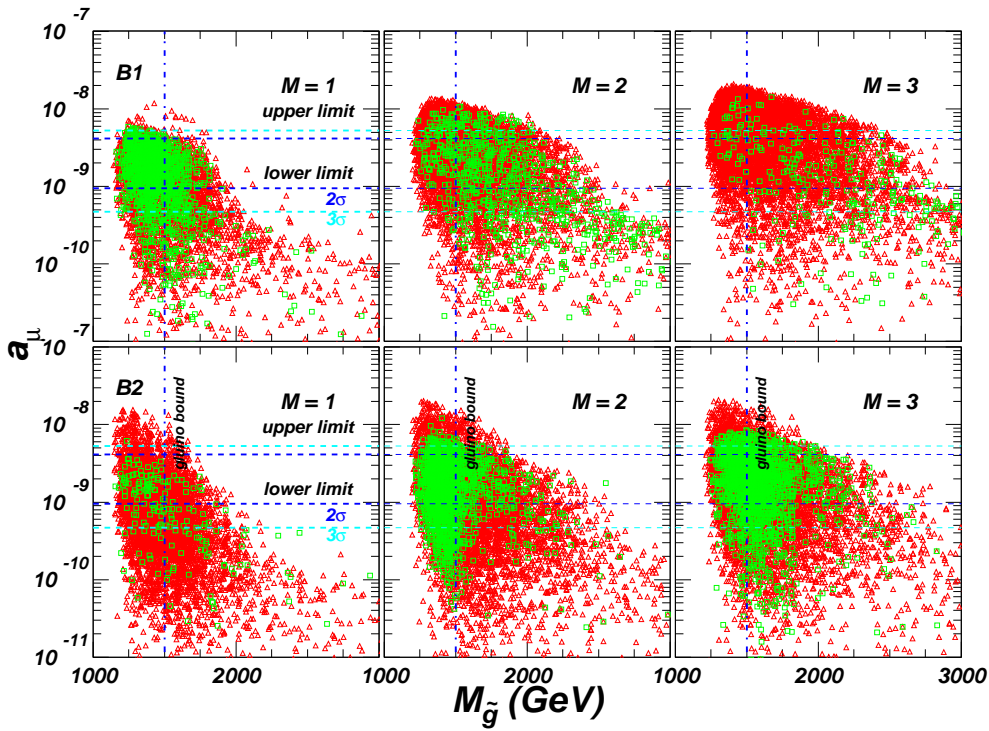}
\end{center}
\vspace{-.8cm}
\caption{The scatter plots of the surived samples showing
 the muon $g-2$ versus the gluino mass in Scenario B with incomplete GUT multiplets (adjoint messengers).
The upper panel corresponds to Scenario B1 while the lower panel is for Scenario B2.
The green $'\Box'$ samples satisfy both the upper and lower bounds of the dark matter relic density.}
\end{figure}

\item  Scenario B:

The scatter plots of the survived samples showing $a_\mu$ versus $m_{\tilde g}$ in Scenario B are shown in Fig.\ref{scB1},
in which the upper panel is for Scenario B1 and the lower pannel is for Scenario B2.
We can see that a lot of points which can fully account for the $g_\mu-2$ anomaly can survive
the LHC gluino mass bound, especially, for a larger $M$.
So scenarios with the incomplete GUT representation of messengers are more favored by the $g_\mu-2$ data.

\begin{figure}[htb]
\begin{center}
\includegraphics[width=5 in]{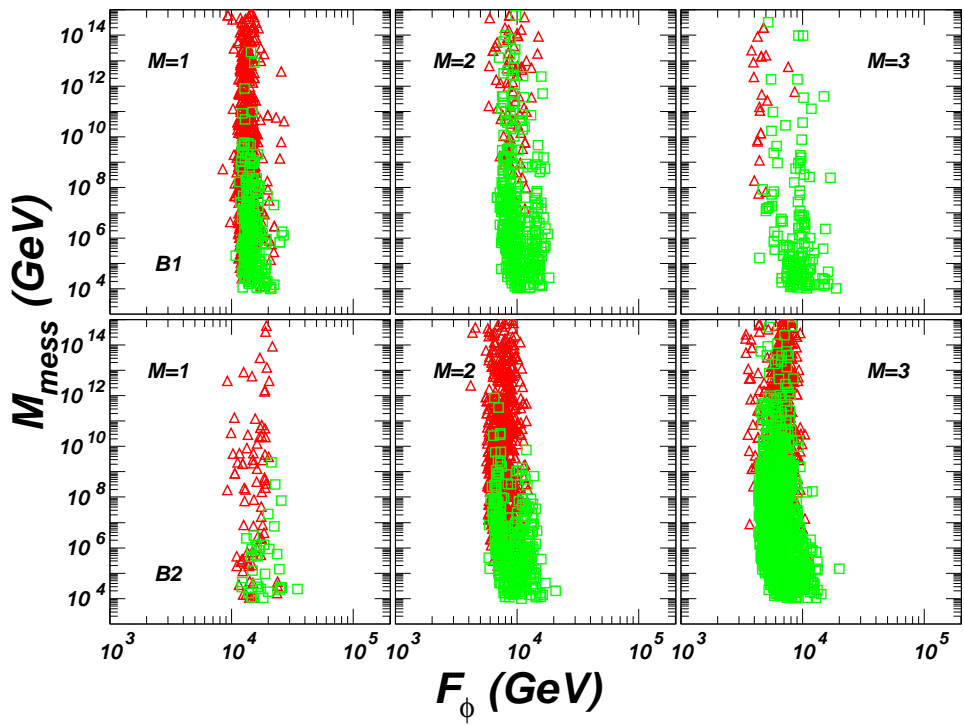}
\end{center}
\vspace{-.8cm}
\caption{ Scatter plots showing $F_\phi$ versus the messenger scale $M_{mess}$ for Scenario B.
All the points satisfy both the lower and upper bounds of dark matter relic density
and collider constraints.
The red $\triangle$ (green $\Box$) samples are excluded (allowed) by the gluino mass bound
$m_{\tl{g}}\geq 1.5$ TeV.
}
\label{scB2}
\end{figure}

Similar to Scenario A, the upper bound of gluino mass can be understood from the upper bound of $F_\phi$,
which is obvious in Fig.\ref{scB2} for both cases.
The upper mass bound of gluino is around 3 TeV (2.7 TeV) in both scenarios if the muon $g-2$ is explained
at $3\sigma$ ($2\sigma$) level.
Such a light gluino will be accessible at future LHC experiments.

\begin{figure}[htb]
\begin{center}
\includegraphics[width=5.5 in]{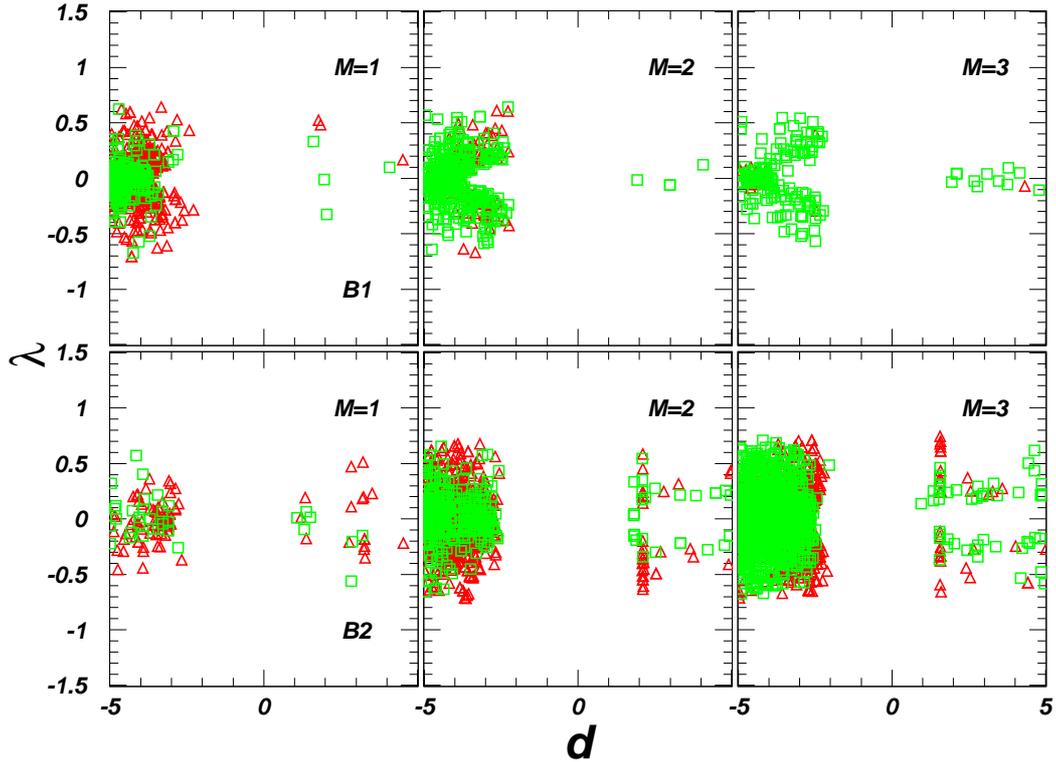}
\end{center}
\vspace{-.8cm}
\caption{Same as Fig.4, but showing the deflection parameter $d$ versus the messenger-matter couplings
$\lambda_E$.
}
\label{scB3}
\end{figure}

The deflection parameter $d$ versus the messenger-matter couplings $\lambda_E\equiv \lambda$
in Scenario B is plotted in Fig.\ref{scB3} with all points satisfying both the upper and lower bound of DM relic density.
 Again, additional non-trivial messenger-matter interactions are obviously advantageous in solving the
$g_\mu-2$ anomaly  with which the allowed range for $d$ enlarges. Besides, the non-vanishing
messenger-matter interactions $\la\neq 0$ can be used to solved the $g_\mu-2$ anomaly for a relatively
small deflection parameter $d$, especially for the Scenario B1. We can see from Fig.\ref{scB3} that
in Scenario B1 the maximum negative $d$ is $-3.5$ with $\la=0$. However, the maximum negative $d$
changes to almost $-2$ with non-vanishing messenger-matter interactions. A small deflection parameter
$|d|$ is relatively easy for model buildings. In Scenario B2, it is not possible to solve the
$g_\mu-2$ anomaly with $\la=0$ for a positive $d$. With messenger-matter interactions, a positive
deflection parameter also works.

In Fig.\ref{scB3} the survived points which satisfy both the upper and lower bounds of dark matter
relic density are shown as green $'\Box'$.
The numerical calculation indicates that the number of points which satisfy the dark matter relic density
decreases with $M$ in Scenario B1, but increases with $M$ in Scenario B2. This can be understood from
the mass ratio between the bino and the gluino with (the most favorite) large negative deflection parameter
$d\sim -4$. For a gluino mass between 1.5 TeV and 3 TeV, the mass ratio should be adjusted to a proper value
at $M_3:M_1\sim {\cal O}(10)$ to fully account for the dark matter relic density by decreasing (Scenario B1)
or increasing (Scenario B2) the value of $M$.
Bino dominated neutralino often leads to over-abundance of DM, unless (co)annihilation processes reduce the relic density to levels compatible with Planck.

We should note that some portion of the parameter space with insufficient DM relic abundance is not
displayed in Fig.\ref{scB2} and Fig.\ref{scB3}. Following the discussions in Scenario A, we obtain
Table 2 from Eq.(\ref{ratio}), showing the range of the deflection parameter $d$ within which the
wino is lighter than bino. Constrained by $F_\phi$, a light wino-like DM will always lead to
insufficient relic abundance.

\begin{figure}[htb]
\begin{center}
\includegraphics[width=5.5 in]{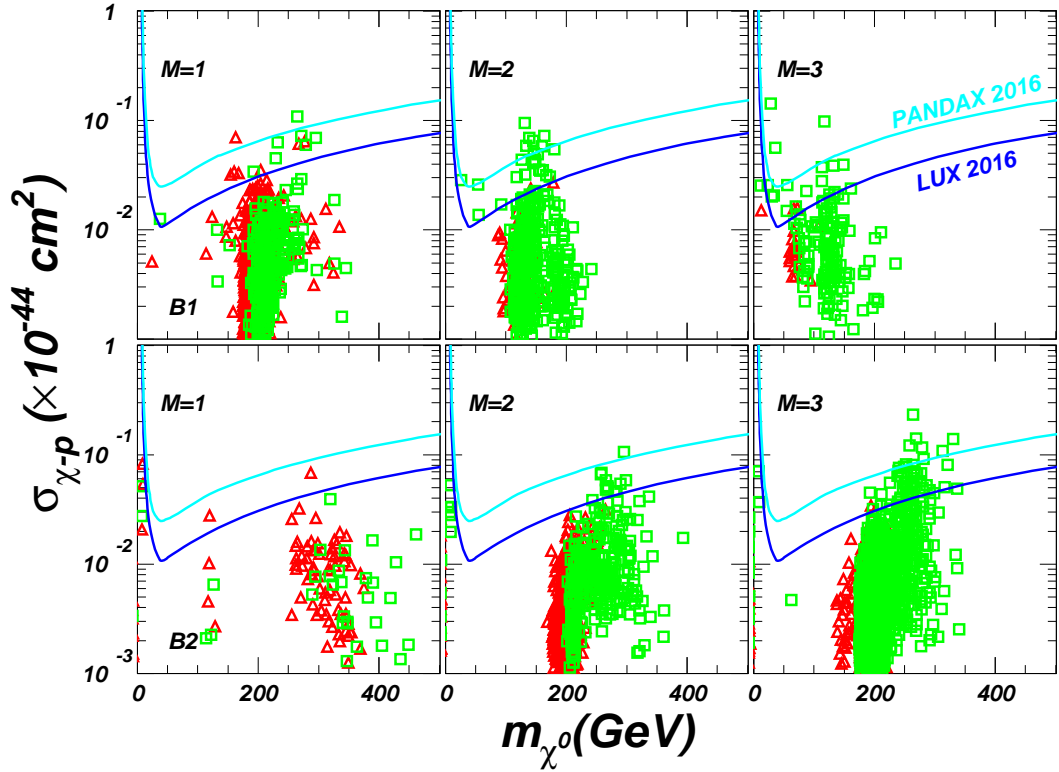}
\end{center}
\vspace{-.8cm}
\caption{Same as Fig.4, but showing the Spin-Independent DM direct detection constraints from LUX 2016 and PandaX.
}
\label{scB4}
\end{figure}

\begin{table}[h]
 \caption{The range of $d$ within which the wino will be lighter than bino for various messenger species $M$
          in Scenario B.}
\centering
\begin{tabular}{|c|c|c|c|c|}
\hline
 &~~~~~~~~&~M=1&~M=2&~M=3\\
\hline
Scenario B1&d&$-0.92\lesssim d\lesssim1.23$ &$-0.51\lesssim d\lesssim 0.78$&$-0.35\lesssim d\lesssim 0.95$\\
\hline
Scenario B2&d&$-1.21\lesssim d\lesssim 1.05$ &$-0.69\lesssim d\lesssim 0.64$&$-0.49\lesssim d\lesssim 0.46$\\
\hline
\end{tabular}
\end{table}
 The DM Spin-Independent(SI) direct detection constraints from LUX and PandaX are shown in Fig.\ref{scB4}. It can be seen that a large portion of points that satisfy the DM relic density can survive the SI direct detection constraints. We know that interactions between bino DM and the nucleons are primarily mediated by t-channel scalar Higgses ($h_0$ and $H_0$), or by s-channel squarks (with t-channel Z-boson exchange process highly suppressed). As the squarks are not found at the LHC, their masses should be significantly larger than the Higgs masses. So the SI cross section is dominated by Higgs-mediated process, despite the associated suppression by Yukawa couplings and the small Higgsino fraction. In scenario B, the type of the neutralino which can give the right DM relic abundance is almost bino-like with small Higgsino component, thus suppress the SI direct detection cross sections.

\eit

\section{\label{sec-3} Conclusions}
We proposed to introduce general messenger-matter interactions
in the deflected anomaly mediated SUSY breaking scenario to explain the $g_\mu-2$ anomaly.
Scenarios with complete or incomplete GUT multiplet messengers are discussed, respectively. The introduction of incomplete GUT mulitiplets can be advantageous in various aspects. We found that the $g_\mu-2$ anomaly can be solved in both scenarios under current
constraints including the gluino mass bounds, while the scenarios with incomplete GUT representation
messengers are more favored by the $g_\mu-2$ data.
We also found that the gluino is upper bounded by about 2.5 TeV (2.0 TeV)
in Scenario A and 3.0 TeV (2.7 TeV) in Scenario B if the generalized deflected AMSB scenarios are used
to fully account for the $g_\mu-2$ anomaly at $3\sigma$ ($2\sigma$) level.
Such a gluino should be accessible in the future LHC searches. 
 Dark matter constraints, including DM relic density and direct detection bounds, favor the scenario B with incomplete GUT multiplets. Much of the allowed parameter space for the scenario B could be covered by the future DM direct detection experiments. 

\section*{Acknowledgement}
This work was supported by the Natural Science Foundation of China under grant
numbers 11375001, 11675147, 11675242, by the Open Project Program of State Key Laboratory of
Theoretical Physics, ITP, CAS (No.Y5KF121CJ1),
by the Innovation Talent project of Henan Province under grant number 15HASTIT017 and the
Young-Talent Foundation of Zhengzhou University, by the CAS Center for Excellence in Particle
Physics (CCEPP), and by the CAS Key Research Program of Frontier Sciences.

\end{document}